\documentclass{article} %
\usepackage{preprint,times}

\usepackage[colorlinks=false,pdfborder={0 0 0}]{hyperref}
\usepackage{url}
\usepackage{graphicx}
\usepackage{float}
\usepackage{amsmath}
\usepackage{amssymb}
\usepackage{xcolor}
\newcommand{\revised}[1]{#1}
\makeatletter
\newcommand{\citepnamebreaksafe}[1]{\begingroup\def\NAT@nmfmt##1{\mbox{\NAT@up##1}}\citep{#1}\endgroup}
\makeatother

\title{Intrinsic timescale placement shapes temporal computation and learning in recurrent networks}

\author{Subati Abulikemu, John Suckling, Michail Mamalakis \\ University of Cambridge \\ \texttt{\{ss2905,js369,mm2703\}@cam.ac.uk}}
\hypersetup{pdftitle={Intrinsic timescale placement shapes temporal computation and learning in recurrent networks},pdfauthor={Subati Abulikemu, John Suckling, Michail Mamalakis}}
\begin{document}
\raggedbottom
\setlength{\parskip}{4pt}

\maketitle

\begin{abstract}

Neural systems comprise units with varying intrinsic timescales, but how this heterogeneity contributes to temporal computation and interacts with recurrent dynamics is not fully resolved. We show that intrinsic heterogeneity provides a richer temporal basis for a readout; however, in our temporal tasks, recurrence can largely reduce this advantage as diverse effective dynamics emerge. By fixing the recurrent graph, its weights, and the set of intrinsic timescales while varying only their assignment across nodes, we find that placement systematically changes the effective temporal repertoire. While degree and strength effects are topology dependent, alignment with slow recurrent modes shows more consistent effects across the tested topology families, indicating slow recurrent structure as a more general organizing coordinate. These dynamical effects also influence function: under fixed recurrent connectivity, slow mode alignment improves temporal computation in the tested tasks and topology families, with generally larger benefits at greater temporal demand, and serves as an inductive bias when recurrence is trainable. \revised{Finally, we examine a corresponding association in human cortical activity. Empirical magnetoencephalography (MEG) activity timescales are positively associated with slow modes of the structural connectome, and their observed spatial ordering produces a broader effective temporal repertoire than spatially rotated alternatives when imposed as model time constants.} Taken together, these results suggest that placement relative to slow recurrent structure is an organizing factor in how intrinsic timescale heterogeneity contributes to network dynamics, computation, and learning.

\end{abstract}

\section{Introduction}
\label{sec:intro}

Neural computation unfolds over timescales from rapid sensory processing to maintenance of information in memory. Neural activity exhibits temporal heterogeneity throughout the cortex and within local populations; sensory regions fluctuate faster than higher-order prefrontal systems, and neurons within a cortical region can display a range of intrinsic timescales \citep{Murray2014,Bernacchia2011,Wasmuht2018,Abulikemu2026}. Such variation in intrinsic timescales is functionally important, with longer timescales associated with stronger and more persistent task representations \citep{Cavanagh2016,Wasmuht2018}.

One source of this variation is heterogeneity in local neuronal and synaptic time constants, which can supply complementary fast and slow temporal filters. Distinct or adaptive time constants are shown to support temporal hierarchy in recurrent networks \citep{Yamashita2008,Quax2020}, and heterogeneous membrane and synaptic time constants improve learning on temporally structured tasks \citep{PerezNieves2021}. Effective empirical activity dynamics, however, do not necessarily correspond to a single local time constant \citep{Zeraati2026}.

Recurrence provides another source of network temporal structure. Recurrent state retains information of past inputs decodable by downstream readouts \citep{Ganguli2008,Dambre2012}, and recurrent interactions modulate the expression of local temporal processes at the network level \citep{Beiran2019}. In linear recurrent systems, eigenvalues determine the decay timescales of dynamical modes, and eigenvectors determine where those modes are spatially expressed in the network \citep{Chaudhuri2014}. Recurrent organization can correspondingly generate distinct and hierarchical activity timescales among assemblies and cortical regions \citepnamebreaksafe{Chaudhuri2015,Stern2023,Zeraati2023}. When local time constants and recurrent weights are jointly optimized, different training regimes can lead to local or network-mediated mechanisms for maintaining long timescales \citep{Khajehabdollahi2024}. The computational contribution of a local time constant therefore depends on its recurrent context. 

\revised{In this context, the contribution of intrinsic heterogeneity to network dynamics cannot be analysed solely from the distribution of local time constants. Recurrent connectivity can already produce diverse effective timescales even with homogeneous local time constants, while the contribution of heterogeneous time constants may depend on which neurons receive the longer or shorter values. Separating the effects of timescale placement from those of heterogeneity therefore allows us to examine how placement affects network dynamics and computation, and whether the initial arrangement influences learning when recurrent weights are trained.} Previous architectures have introduced temporal hierarchy by assigning different update rates to modules or different time constants to layers \citep{Koutnik2014,Moro2026}. Here, we directly examine timescale placement by altering which nodes receive which time constants and keeping the recurrent network and timescale set fixed. In-strength and degree are candidate structural coordinates for placement, as greater incoming connectivity is linked with slower dynamics \citep{Sethi2017}, and degree heterogeneity can facilitate slow hub dynamics \citep{Zenari2026}. These measures, however, summarize local connectivity, whereas recurrent modes provide a network-level summary of collective decay dynamics \citep{Chaudhuri2014,Tang2020}. We therefore hypothesize that aligning slower intrinsic timescales with stronger participation in slow recurrent modes provides a more direct placement coordinate. We compute this coordinate independently of the tested assignment, through a reference network whose units all share the same local time constant.

For such organization to be computationally meaningful, placement should have effects beyond spontaneous dynamics, and its effect should become more apparent when information must be retained for longer periods. \revised{We use memory and decision tasks with varying delays to examine how timescale placement affects computation under fixed recurrent weights, and whether the initial placement also influences learning when recurrent weights are trained.} We study this functional effect by comparing task performance across placements with fixed recurrent weights, then enable recurrence to adapt to test whether aligned placement provides an advantageous starting organization. Pre-learning dynamics can influence learning speed \citep{Kurikawa2025}, and recurrent learning can develop longer network-mediated dynamics without corresponding changes in local time constants \citep{Khajehabdollahi2024}. Thus, placement may act as a dynamical inductive bias, and recurrent plasticity may partly compensate for the initially unfavorable placement.  

A complementary question is whether human cortical organization parallels this placement hypothesis. Resting functional MRI timescales are associated with structural connectivity strength \citep{Fallon2020}, and connectome-constrained models show that white matter connectivity shapes regional temporal processing \citep{Triebkorn2025}. \revised{Recent network control modeling links inferred heterogeneous regional decay rates to empirical activity timescales and changes in connectome eigenmodes} \citep{Kim2025}. Here, we investigate whether an independently measured cortical timescale map aligns with slow modes of the structural connectome and whether its empirical spatial arrangement leads to predicted dynamical consequences. We use intrinsic activity timescales estimated from magnetoencephalography (MEG) \citep{Shafiei2023}, which are treated as regional neural activity measures rather than direct measurements of local time constants \citep{Zeraati2026}.

\textbf{Contributions} We present four main findings. \revised{First, we isolate timescale placement from the recurrent network and timescale distribution, and show that slow mode alignment organizes the realized temporal repertoire more consistently than local structural coordinates and weighted eigenvector centrality}. Second, we establish functional benefit of this alignment for temporal computation with increasing temporal demand. Third, we show that placement biases persist under trainable recurrence, while recurrent plasticity partly compensates for unfavorable placement and reorganizes slow recurrent structure. Lastly, we \revised{find that measured cortical activity timescales are associated with slow structural connectome modes, and that their spatial ordering broadens the effective temporal repertoire in the connectome model}.

\section{Methods}
\label{sec:methods}

\subsection{Experimental overview}
\label{sec:method_overview}

\begin{figure}[!htbp]
    \centering
    \includegraphics[
        width=\textwidth,
        trim=79bp 16bp 75bp 10bp,
        clip
    ]{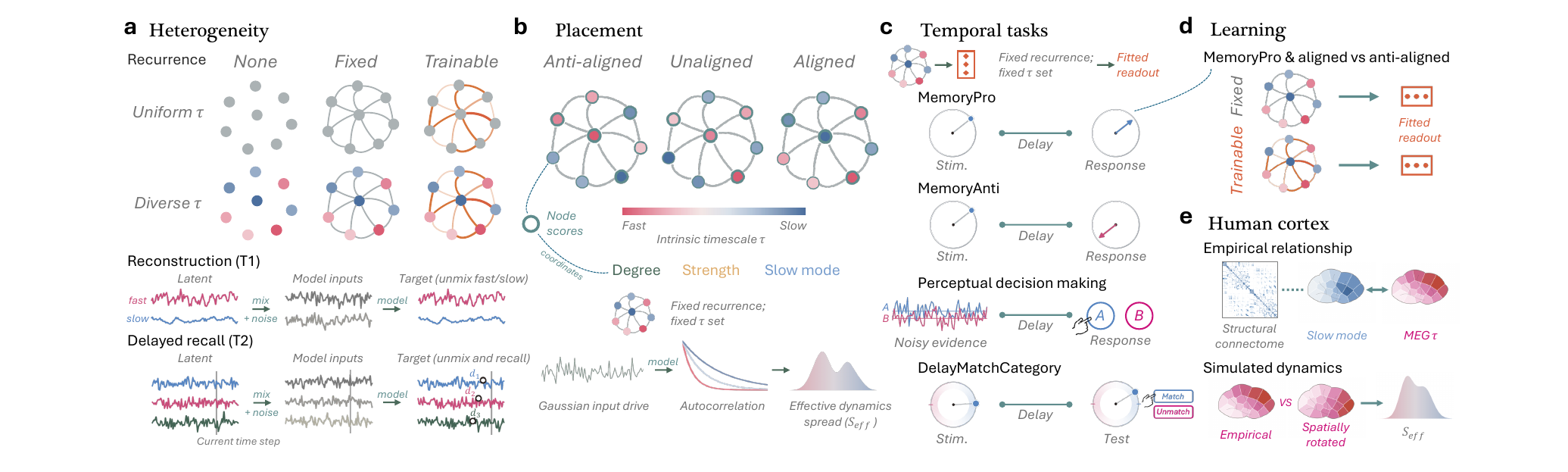}
    \caption{\textbf{Experimental overview.}
    \textbf{a} Intrinsic timescale heterogeneity is compared under no, fixed, and trainable recurrence with two diagnostic tasks, latent reconstruction (T1) and delayed recall (T2).
    \textbf{b} Keeping recurrent graph, weights, and timescale set fixed, timescales are permuted according to degree, strength, or slow mode coordinates. A Gaussian drive is used to estimate effective timescales and their spread \(S_{\mathrm{eff}}\).
    \textbf{c} Four cognitive tasks test functional consequences of placement with fixed recurrence and fitted readouts.
    \textbf{d} Learning is assessed using MemoryPro with aligned and anti-aligned placements under frozen or trainable recurrence; readouts fitted in both.
    \textbf{e} Human cortical analyses link empirical MEG timescales to slow connectome structure and compare the effective temporal repertoire in empirical and spatially rotated placements. Task signals, distributions, and cortical maps are shown schematically.}
    \label{fig:pipeline}
\end{figure}

Figure~\ref{fig:pipeline} summarizes the five-stage design used in this paper. \revised{First, we examine the contribution of intrinsic timescale heterogeneity to network computation, with and without recurrence (Result~\ref{sec:result1}). Second, we investigate whether changing the assignment of intrinsic time constants produces different effective dynamics while keeping recurrent weights and the set of time constants fixed (Result~\ref{sec:result2}). Third, we assess how timescale placement affects performance on memory and decision tasks (Result~\ref{sec:result5}). Fourth, we explore the effect of placement on early learning when recurrent weights are trainable (Result~\ref{sec:result7}). Finally, we compare measured cortical activity timescales with slow modes of the human structural connectome (Result~\ref{sec:result_connectome}).}

In model placement experiments, placement rules are defined from recurrent network properties before any task performance is evaluated, and effects are estimated by comparing different timescale assignments within the same independently generated network. The initial timescale heterogeneity and untasked placement experiments use 64-unit networks; the fixed recurrence behavioral study and recurrent learning study use 128-unit Erdős-Rényi (ER), heavy-tail, and modular networks. The cortical analysis uses a fixed 400-parcel human structural connectome with empirical MEG activity timescales mapped to the corresponding parcels.

\subsection{Network model and temporal organization}
\label{sec:method_temporal}

We study leaky recurrent neural networks (RNNs) with neuron-level intrinsic timescales,
\begin{equation}
 h_i(t)=(1-\alpha_i)h_i(t-1)+\alpha_i\,\phi\!\left((W_{\mathrm{in}}x_t)_i+(W_{\mathrm{rec}}h_{t-1})_i+b_i\right),
 \qquad \alpha_i=1-e^{-\Delta t/\tau_i},
\end{equation}
where \(h_i(t)\) is the state of unit \(i\), \(x_t\) is the input, and \(\phi\) is ReLU; larger \(\tau_i\) results in slower updating. The timescale heterogeneity experiments compare non-recurrent leaky units (\(W_{\mathrm{rec}}=0\)), a reservoir with fixed recurrence, and an RNN with trainable recurrence (Fig.~\ref{fig:pipeline}a). Homogeneous models learn one globally shared \(\tau\), and heterogeneous models learn separate \(\tau_i\) for each unit. We use two diagnostic tasks; T1 recovers latent signals evolving at different speeds from noisy mixtures, and T2 recalls random latent inputs after fixed delays. Performance \(G\) is normalized relative to a memoryless baseline (\(G=0\)) and a task-specific analytic reference (\(G=1\)) (Appendix~\ref{app:tasks}).

Recurrence can make a unit's activity timescale differ from its intrinsic \(\tau_i\). Under a common Gaussian white noise drive, we estimate the effective timescale \(\tau_i^{\mathrm{net}}\) from the unit autocorrelation and define the realized temporal repertoire as \(S_{\mathrm{eff}}=\operatorname{std}_i[\log\tau_i^{\mathrm{net}}]\), using logarithms to capture proportional differences between timescales. Larger \(S_{\mathrm{eff}}\) denotes a broader spread of effective timescales, which reflect the interaction between intrinsic and recurrent dynamics. Across placement studies, we use structural coordinates including degree and absolute weighted in-strength, and a spectral slow mode coordinate. To characterize where the recurrent network itself supports slow dynamics independently of heterogeneous intrinsic timescales, and motivated by eigenvector localization \citep{Chaudhuri2014}, we define the slow mode coordinate from a homogeneous reference system,
\begin{equation}
J_0=(1-\alpha_0)I+\alpha_0\bar{D}W_{\mathrm{rec}},
\end{equation}
where \(\alpha_0=1-e^{-\Delta t/\tau_0}\) corresponds to a single reference timescale \(\tau_0\), and \(\bar{D}_{ii}\) is the proportion of timesteps for which unit \(i\) has a nonzero ReLU activation under a fixed Gaussian reference drive. Each node is scored by its loading onto the modes of \(J_0\), where greater weights are given to modes decaying more slowly, so larger values indicate stronger participation in slow recurrent activity. As the reference system and resulting coordinate are computed before heterogeneous timescales are assigned, they do not depend on the placement being evaluated (Appendix~\ref{app:slowmode}). The connectome analysis uses the same slow mode construction with a linear reference system defined in Section~\ref{sec:method_connectome}.

\subsection{Timescale placement and temporal computation with fixed recurrence}
\label{sec:method_fixed}

In the untasked placement study, we vary the arrangement of intrinsic timescales using the structural and slow mode coordinates defined in the previous section and measure the resulting effective dynamics (Fig.~\ref{fig:pipeline}b). We use 64-unit reservoirs and permute timescales to vary the Spearman correlation between \(\tau_i\) and degree, in-strength, or the slow mode coordinate from anti-aligned to aligned configurations. All placements receive the same Gaussian drive, and the resulting temporal repertoire is quantified by \(S_{\mathrm{eff}}\) (Appendices~\ref{app:placement}--\ref{app:slowmode}).

To test whether the same placement intervention influences temporal computation, we use 128-unit ER, heavy-tail, and modular reservoirs with \(\Delta t=20\) ms and a shared set of 128 log-spaced timescales from 40 to 640 ms. Slow mode placement is the primary manipulation, with degree placement serving as a structural comparison. We evaluate 16 independent networks per topology at seven target alignment levels from \(-1\) to \(1\). We also test the primary placement effect for size robustness in 64-, 128-, and 256-unit networks (Appendix~\ref{app:taskedplacement}).

We vary temporal demand across seven delay levels (up to 2400 ms) in four cognitive tasks (Fig.~\ref{fig:pipeline}c). MemoryPro requires maintaining a stimulus direction across a delay and responding in the same direction, whereas MemoryAnti requires responding in the opposite direction \citep{Yang2019,Driscoll2024}. The perceptual decision making (PDM) task requires integrating noisy stimulus evidence into a binary decision and retaining it through a delay \citep{Yang2019,Inagaki2019}. DelayMatchCategory requires retaining the category of a sample stimulus and deciding whether a later test stimulus belongs to the same category \citep{FreedmanAssad2006}. For each placement, we fix the reservoir and fit a ridge-regularized linear readout.

The ridge penalty, task difficulty, and intrinsic timescale range are calibrated using strictly random placements without inspecting aligned versus anti-aligned performance. The primary behavioral measure is the error contrast \(B(d)=E_{\mathrm{anti}}(d)-E_{\mathrm{aligned}}(d)\), where positive values indicate an aligned placement advantage. At 2400 ms, we estimate 95\% confidence intervals using paired bootstrap resampling across networks and assess statistical significance with paired sign-flip tests. As a secondary analysis, we examine the relationship between placement-induced \(S_{\mathrm{eff}}\) and behavior. \revised{We use weighted eigenvector centrality based on absolute recurrent weights as an additional placement baseline and compare placements with similar mean log effective timescales but different spreads. We also perturb slow and fast recurrent modes during memory delays, adjusting their strengths to disrupt similar amounts of network activity using task-independent Gaussian inputs (Appendix~\ref{app:additional_controls}).}

\subsection{Learning and recurrent plasticity}
\label{sec:method_learning}

To test whether placement acts as an inductive bias when recurrence can adapt, we use 24 new 128-unit RNNs (eight per topology family) with MemoryPro at delays of 400, 800, 1600, and 2400 ms. Intrinsic timescales are placed at moderate target correlations of \(\pm0.5\) with either the slow mode coordinate or degree; slow mode placements are constrained to remain nearly neutral with respect to degree. Each placement is evaluated with recurrence either frozen or trainable, with the output readout trained in both conditions. Paired frozen and trainable runs use matched initialization, training data, and evaluation trials (Appendix~\ref{app:learning}).

\revised{We measure early learning as the cumulative improvement in held-out error relative to the initial error of each run. For networks with trainable recurrent weights, we examine how the learning advantage of aligned versus anti-aligned placement changes with memory delay, and whether slow mode alignment predicts this effect after accounting for degree alignment.}

\revised{At 2400 ms, we compare the late performance gap between aligned and anti-aligned networks with fixed and trainable recurrent weights to assess how much recurrent training compensates for unfavorable placement. Since intrinsic time constants remain fixed during training, changes in their alignment with slow recurrent modes arise from changes in recurrent weights. We track this alignment in initially anti-aligned networks and use network-level bootstrap and paired tests (Appendix~\ref{app:learning}).} \revised{In a separate MemoryPro experiment, we compare fixed and learned intrinsic time constants in nine networks with either fixed or trainable recurrent weights, and examine whether the learned time constants align with slow recurrent modes (Appendix~\ref{app:learnable_tau}).}

\subsection{Human connectome and cortical timescale organization}
\label{sec:method_connectome}

The final analysis examines the consistency of human cortical timescale organization with slow mode placement and its predicted dynamics. We use a group-representative structural connectome from 327 Human Connectome Project (HCP) participants \citep{VanEssen2013,Milisav2026} and a group-average MEG timescale map from 33 HCP participants \citep{Shafiei2023}, both in the 400-region Schaefer atlas \citep{Schaefer2018}. The structural connectivity matrix \(A\) is spectrally normalized as \(W=gA/\lambda_{\max}(A)\), with near-critical gain \(g=0.9\). We define the homogeneous linear reference system
\begin{equation}
J_0=(1-\alpha_0)I+\alpha_0W,
\end{equation}
with \(\alpha_0=1-e^{-1/\tau_0}\) and no ReLU gating. Given that \(J_0\) and \(W\) share eigenvectors, each parcel is mapped by its participation in connectome eigenmodes with slower modes weighted more strongly. We test the association between cortical timescales and slow mode score using Spearman correlation.

To test the dynamical consequence, the structural connectome and timescale set are fixed, and spatially rotated MEG maps provide null placements that preserve cortical organization but change their relation to slow mode structure. For each placement, the linear reservoir with recurrence \(W\) is driven by Gaussian white noise and its temporal repertoire quantified with \(S_{\mathrm{eff}}\). The empirical ordering is compared with the rotated null using a one-sided permutation test. \revised{We also assign identical intrinsic time constants to all connectome regions and compare the effective timescales generated by this model with MEG data. We further test whether slow mode participation provides additional predictive information beyond the homogeneous model (Appendix~\ref{app:connectome}).}

\section{Results}
\label{sec:results}

\subsection{Intrinsic heterogeneity enriches temporal representations; recurrence reduces its advantage}
\label{sec:result1}

\begin{figure}[!t]
    \centering
    \includegraphics[
        width=\textwidth,
        trim=19 33 31 20,
        clip
    ]{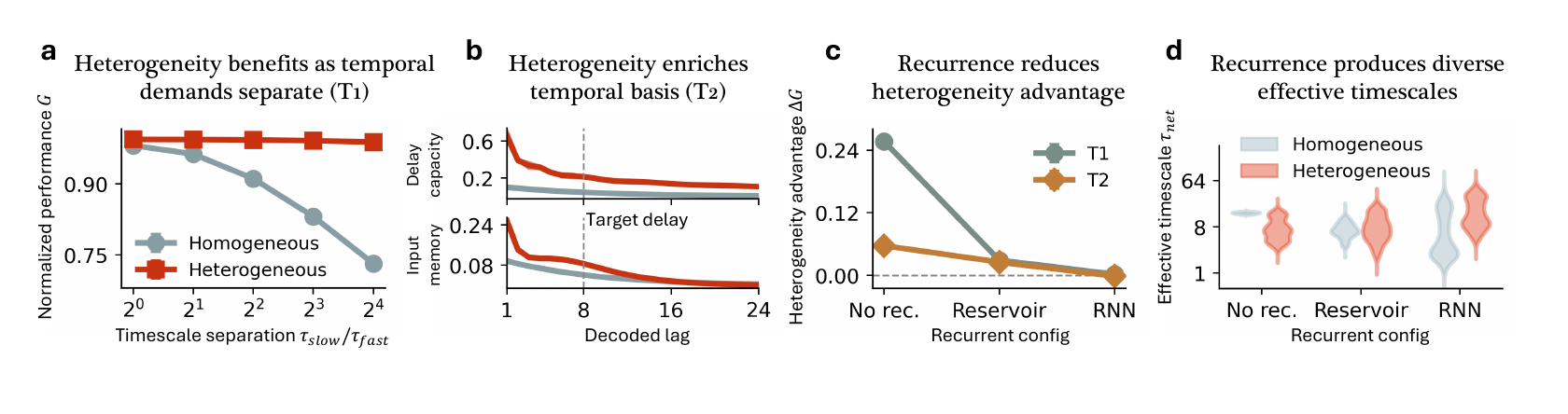}
    \caption{\textbf{Intrinsic heterogeneity enriches temporal representations and recurrence reduces its advantage in the diagnostic tasks.}
    \textbf{a} In non-recurrent leaky units, heterogeneous networks (red) exceed homogeneous networks (gray) on T1, and the difference increases with latent timescale separation.
    \textbf{b} In T2, heterogeneity provides a richer set of temporal kernels (top) and better recovery of past input (bottom); dashed line represents the target delay. Definitions of both diagnostics are given in Appendix~\ref{app:tasks}.
    \textbf{c} The heterogeneity advantage \(\Delta G\) declines as recurrence is added on both tasks.
    \textbf{d} With recurrence, homogeneous networks develop a spread of \(\tau_{\mathrm{net}}\) approaching that of heterogeneous networks.}
    \label{fig:main}
\end{figure}

\revised{The first question is whether intrinsic heterogeneity provides useful temporal filters and how recurrence changes its contribution.} Without recurrence, intrinsic timescale heterogeneity improved temporal representations, particularly when the underlying temporal structure was more diverse. On T1, the heterogeneous advantage increased from \(\Delta G=0.013\) when the two latent signals had the same timescale to \(0.256\) when their timescales separated by a factor of \(2^4\) (Fig.~\ref{fig:main}a). T2 showed a complementary view of this benefit (Fig.~\ref{fig:main}b); heterogeneous timescales supplied distinct temporal kernels that a readout can combine, resulting in better representation of a delayed target (\(0.214\) versus \(0.043\) at the delay target 8) and more recoverable past input (\(1.388\) versus \(0.816\) summed over lags).

This intrinsic advantage diminished once recurrence could itself generate temporal diversity. Across non-recurrence, reservoir, and trained RNN, homogeneous performance on the hardest T1 condition increased from \(G=0.731\) to \(0.934\) to \(0.994\), while heterogeneous performance remained high throughout. The advantage decreased from \(0.257\)~\text{95\% CI}~\([0.252,0.260]\) to \(0.029\,[0.022,0.039]\) and \(0.002\,[0.001,0.003]\), and T2 displayed the same effect (Fig.~\ref{fig:main}c). Moreover, homogeneous networks developed a spread of effective timescales despite sharing a single intrinsic timescale, with \(S_{\mathrm{eff}}\) increasing from \(0.046\,[0.042,0.053]\) to \(0.296\,[0.237,0.381]\) and \(1.076\,[0.856,1.251]\) (Fig.~\ref{fig:main}d). Thus, in these task settings, recurrence can largely supply the effective temporal heterogeneity that intrinsic diversity otherwise provides. This shifts the focus from whether heterogeneous timescales are present to how a fixed set of intrinsic timescales is organized relative to recurrent structure.

\raggedbottom
\subsection{Timescale placement shapes effective dynamics through slow modes}
\label{sec:result2}

\begin{figure}[!b]
    \centering
    \includegraphics[
        width=\textwidth,
        trim=63 34 29 24,
        clip
    ]{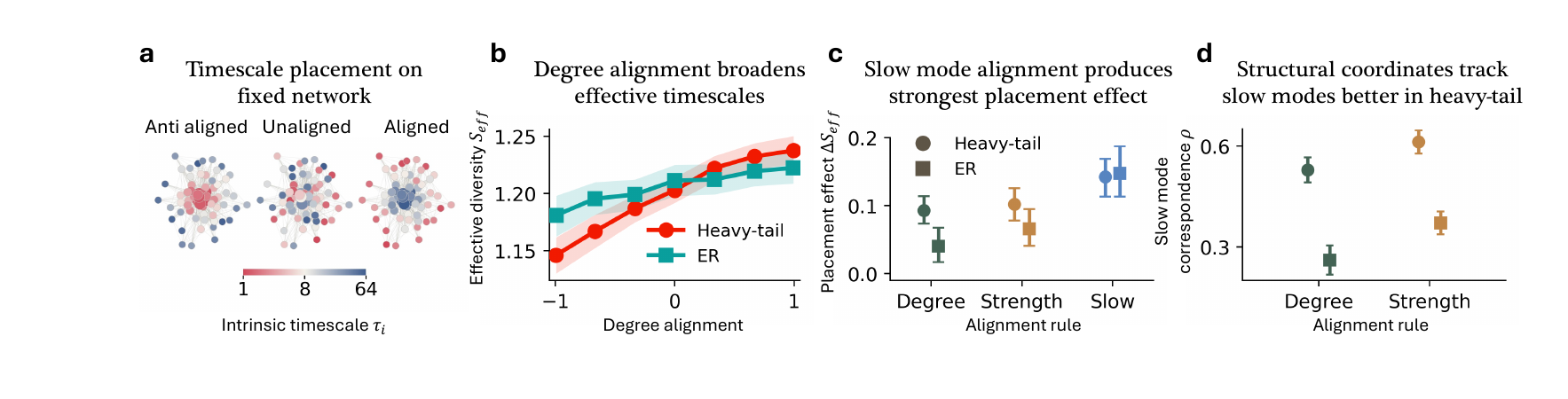}
    \caption{\textbf{Timescale placement and alignment coordinate shape the effective temporal repertoire.}
    \textbf{a} The same reservoir and intrinsic timescale set under different nodewise \(\tau_i\) assignments; colors show intrinsic timescales.
    \textbf{b} Effective diversity \(S_{\mathrm{eff}}\) increases as slower timescales move onto higher degree nodes, with a stronger effect in heavy-tail than ER networks.
    \textbf{c} Across placement coordinates, degree and strength show smaller effects in ER, whereas slow mode alignment shows the largest effect and is similar between topologies.
    \textbf{d} Spearman correlation between structural coordinates and the slow mode coordinate is stronger in heavy-tail networks. \revised{The slow mode coordinate is computed from the homogeneous reference network, independently of the tested intrinsic timescale assignments.}}
    \label{fig:placement_alignment}
\end{figure}

\revised{Recurrence can produce a wide variety of effective timescales, but how these timescales are expressed may depend on the assignment of intrinsic time constants across neurons.} The realized temporal repertoire changed when the same intrinsic timescales were permuted across a fixed reservoir. Here, the graph structure, its connection weights, and the complete set of intrinsic timescales were retained, and only which node received which timescale was varied (Fig.~\ref{fig:placement_alignment}a). Placing slow timescales onto high-degree nodes broadened \(S_{\mathrm{eff}}\), with within-network slopes of \(0.048\,[0.038,0.058]\) in heavy-tail networks and \(0.019\,[0.008,0.032]\) in ER networks (Fig.~\ref{fig:placement_alignment}b). When comparing extremes, the aligned regime exceeded the anti-aligned regime by \(0.092\,[0.071,0.112]\) in heavy-tail versus \(0.041\,[0.017,0.070]\) in ER. Thus, identical intrinsic timescales can generate different effective dynamics depending on their arrangement, with a larger degree-based effect when node degree is more heterogeneous.

The topology dependence of degree points to a dynamical rather than purely structural explanation. Degree describes local network wiring, whereas the slow mode coordinate quantifies each node's loading onto slowly decaying recurrent modes. Among all three placement coordinates, slow mode alignment produced the largest effect in both topologies, increasing \(S_{\mathrm{eff}}\) by \(0.142\,[0.113,0.169]\) in heavy-tail and \(0.147\,[0.113,0.187]\) in ER networks (Fig.~\ref{fig:placement_alignment}c). Degree and strength effects were weaker in ER, in line with their weaker correspondence to the slow mode coordinate.  
Degree correlated more strongly with the slow mode coordinate in heavy-tail than ER networks (\(\rho=0.529\) versus \(0.263\)), and similarly for in-strength (\(0.612\) versus \(0.371\); Fig.~\ref{fig:placement_alignment}d). Structural coordinates therefore manifest stronger placement effects when they track slow recurrent modes better, suggesting that slow mode organization is the more consistent coordinate across the tested topologies.

\revised{We also compared slow mode placement with weighted eigenvector centrality, computed from the absolute recurrent weights of the reservoir. For 24 independently generated 128-unit reservoirs (eight per topology), we kept the recurrent graph, its weights, and the intrinsic timescale values fixed when comparing the two placement rules. Slow mode placement produced a larger increase in effective timescale diversity than weighted centrality (mean difference in $\Delta S_{\mathrm{eff}}=0.0593$, positive in 23 of 24 networks; Appendix~\ref{app:additional_controls}). These results show that the advantage of slow mode placement is not simply due to using recurrent connectivity weights as a placement criterion.}

\flushbottom
\subsection{Slow mode placement shapes temporal computation across demand}
\label{sec:result5}
\begin{figure}[!htbp]
    \centering
    \includegraphics[
        width=\textwidth,
        trim=8 40 38 44,
        clip
    ]{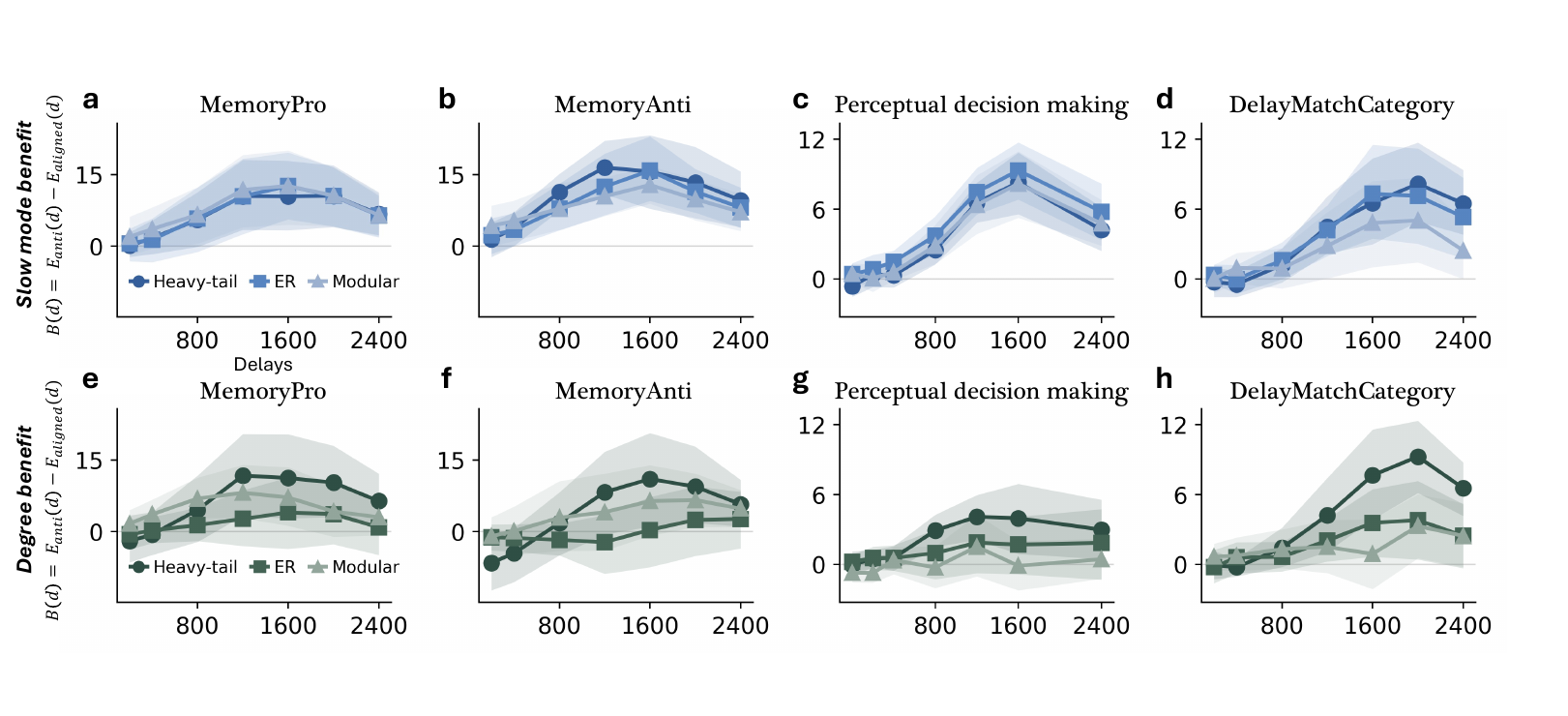}
    \caption{\textbf{Timescale placement shapes temporal computation across demand.}
    Aligned placement benefit \(B(d)=E_{\mathrm{anti}}(d)-E_{\mathrm{aligned}}(d)\) with increasing temporal demand; positive values indicate better performance under aligned regime.
    \textbf{a-d}, slow mode placement for MemoryPro, MemoryAnti, PDM, and DelayMatchCategory, respectively.
    \textbf{e-h}, degree placement for the same four tasks.
    Slow mode benefits are generally larger at higher temporal demand and are coherent in ER, heavy-tail, and modular networks; degree placement effects are strongest in heavy-tail networks and weaker and mixed in the other topologies.
    Curves show topology-specific means across independently generated networks within each topology, and shaded regions show 95\% confidence intervals.}
    \label{fig:tasked_placement}
\end{figure}

\revised{This variation in effective dynamics may affect performance in tasks where information must be maintained over longer periods.} In the reservoir functional experiment, aligned slow mode placement improved performance across all four tasks and all three network topologies at the maximal 2400 ms delay (Fig.~\ref{fig:tasked_placement}a-d). Effects were coherent for MemoryPro (\(d_z=0.70,\,0.70,\,0.69\); all \(p\leq.01\)) and MemoryAnti (\(d_z=0.89,\,0.79,\,0.89\); all \(p\leq.002\)) across ER, heavy-tail, and modular networks, respectively. PDM showed the largest functional effects (\(d_z=1.12,\,1.05,\,1.15\); all \(p<.001\)), and DelayMatchCategory showed variable but positive effects (\(d_z=0.72,\,1.11,\,0.46\); \(p=.007,\,<.001,\,.043\)). The benefit of alignment was generally larger at higher temporal demand. Relative to the lowest demand condition, the aligned benefit was larger at 2400 ms in 9 of 12 task-by-topology comparisons at \(p<.05\). This increase was observed in all three topologies for PDM and in ER and heavy-tail networks for MemoryPro, MemoryAnti, and DelayMatchCategory. Across the full trajectories, effects were generally small at low demand and increased as information had to be maintained over longer delays, peaking around 1600-2000 ms before attenuating at the hardest condition (Fig.~\ref{fig:tasked_placement}a-d). Aligned slow mode placement thus became more influential as temporal demand increased, although the relationship was not strictly monotonic.

Degree placement provided a structural comparison to this pattern, where its effect was strongly topology dependent (Fig.~\ref{fig:tasked_placement}e-h). In heavy-tail networks, all four effects at 2400 ms were positive (\(d_z=0.54\)-\(1.36\), all \(p<.05\)); in ER networks, effects were substantially weaker (\(d_z=0.08\)-\(0.42\), all \(p>.05\)). Modular networks were mixed, with evidence for MemoryAnti (\(d_z=0.58\), \(p=.018\)) but weaker effects in the remaining tasks (\(p=.056\)-\(.316\)). This coincides with the untasked dynamics; degree shows stronger effects where it better tracks slow recurrent modes, whereas slow mode placement shows a more consistent behavioral effect across network architectures.

\revised{We found that at 1600 ms, the difference in PDM error between anti-aligned and aligned placements was 8.34 percentage points for slow mode placement, compared to 3.15 percentage points for weighted centrality (paired difference = 5.19 [2.76, 7.47], $p<.001$). This difference between the two placement rules remained at 2400 ms (2.59 [0.13, 4.91], $p=.003$; Appendix~\ref{app:additional_controls}). Thus, slow mode placement is able to outperform a placement based on recurrent edge weights, when the same recurrent weights, intrinsic timescale values, and PDM settings are used.} \revised{Across seven alignment levels, the correlations between placement-induced $S_{\mathrm{eff}}$ and task benefit generally increased with delay, after controlling for mean effective timescale. For the 12 task-by-topology estimates at 2400 ms, all were positive and 10 of these had confidence intervals that did not include zero (Appendix~\ref{app:taskedplacement}).} 

\revised{We also conducted two functional controls. For 22 of 24 reservoirs, we found that placements with similar mean log effective timescales but greater spread in effective timescales had on average 7.53 percentage points lower PDM error at 1600 ms ([5.48, 9.44], $p<.001$). In the second control, we perturbed the slow recurrent modes during the memory delay and compared the results to those obtained by perturbing the fast modes of comparable strength. We found that the aligned benefit was reduced more when the slow modes were perturbed (6.47 [4.21, 8.82], $p<.001$ at 1600 ms). Both effects remained at 2400 ms (Appendix~\ref{app:additional_controls}).}

\subsection{Timescale placement shapes learning and recurrent plasticity compensates for unfavorable organization}
\label{sec:result7}

\begin{figure}[!htbp]
    \centering
    \includegraphics[
        width=\textwidth,
        trim=17 33 34 21,
        clip
    ]{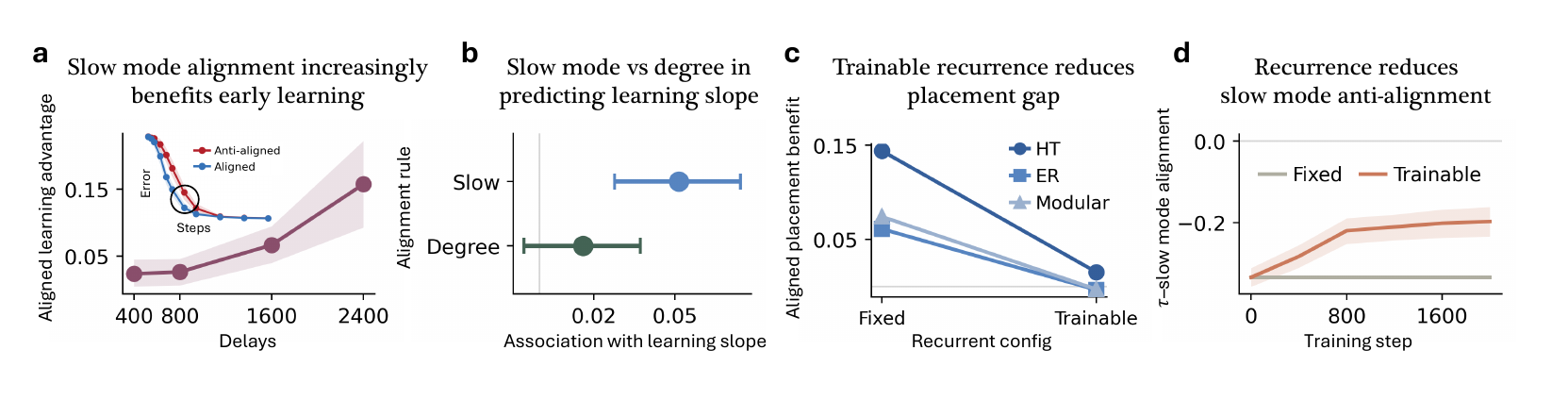}
    \caption{\textbf{Timescale placement shapes learning; recurrence compensates with reorganization.}
    \textbf{a} Aligned slow mode placement improves early learning, with the advantage increasing with temporal demand.
    \textbf{b} Across trainable conditions, slow mode alignment is more strongly associated with the demand dependent learning slope than degree alignment.
    \textbf{c} At 2400 ms, the aligned versus anti-aligned performance gap is smaller with trainable than frozen recurrence.
    \textbf{d} In initially anti-aligned networks, trainable recurrence increases alignment between fixed intrinsic timescales and slow recurrent structure.
    Bars and shaded regions show 95\% confidence intervals.}
    \label{fig:learning_plasticity}
\end{figure}

\revised{The initial placement of intrinsic timescales can also affect learning when recurrent weights are trained.} We observed that aligned slow mode placement accelerated early learning, with its advantage over anti-aligned placement increasing from \(0.023\) at 400 ms to \(0.158\) at 2400 ms (Fig.~\ref{fig:learning_plasticity}a). Within each network, this advantage increased with delay (slope \(=0.048\) \([0.025,0.070]\)); 20 of 24 networks showed a positive slope (\(p<.001\)). Thus, slow mode placement functions as a demand dependent inductive bias, with alignment becoming more beneficial as temporal demand grows.

When all trainable recurrence conditions were considered together, the demand dependent learning effect was more closely associated with slow mode alignment than degree alignment (Fig.~\ref{fig:learning_plasticity}b). After accounting for degree alignment and network identity, slow mode alignment was associated with the demand dependent learning slope (\(\beta_{\mathrm{slow}}=0.051\,[0.028,0.074]\), \(p<.001\)), whereas degree alignment showed a smaller independent association (\(\beta_{\mathrm{degree}}=0.016\,[-0.006,0.037]\), \(p=.147\)). The slow mode coefficient remained positive when each topology family was omitted in turn (Appendix~\ref{app:learning}), indicating that it is a coordinate more coherently associated with the learning benefit.

Recurrent plasticity partly compensated for initial anti-alignment. At the maximal 2400 ms delay, we compared the aligned versus anti-aligned performance gap in matched networks with recurrence either frozen or trainable (Fig.~\ref{fig:learning_plasticity}c). This placement gap was smaller when recurrence was trainable, with a mean reduction of 0.090 [0.046, 0.136]; the reduction was positive in 20 of 24 networks (\(p<.001\)). Intrinsic timescales attached to each neuron remained fixed during training, so any increase in their alignment with slow recurrent structure had to arise from changes in the recurrent weights. In line with this, initially anti-aligned networks developed greater alignment between their fixed intrinsic timescales and the evolving slow recurrent structure under trainable recurrence (Fig.~\ref{fig:learning_plasticity}d). The pre- to post-training increase in slow mode alignment was \(0.137\,[0.108,0.166]\), positive in 23 of 24 networks (\(p<.001\)). Recurrent plasticity therefore reduced the behavioral cost of anti-aligned placement while the slow recurrent structure reorganized toward the fixed intrinsic timescale pattern.

\revised{In nine independent MemoryPro networks, learning intrinsic $\tau$ reduced late normalized angular error with frozen recurrence (mean reduction 0.1365, $p=.004$), but gave little additional improvement when the recurrent weights were trainable (0.00109). Neither condition showed positive mean final alignment with slow recurrent modes (Spearman $\rho=-0.017$ and $-0.027$); task learning therefore does not necessarily converge to aligned placement (Appendix~\ref{app:learnable_tau}).}

\subsection{Human cortical timescale organization aligns with the predicted dynamics of slow mode placement}
\label{sec:result_connectome}

\begin{figure}[!htbp]
    \centering
    \includegraphics[
        width=\textwidth,
        trim=19 53 29 51,
        clip
    ]{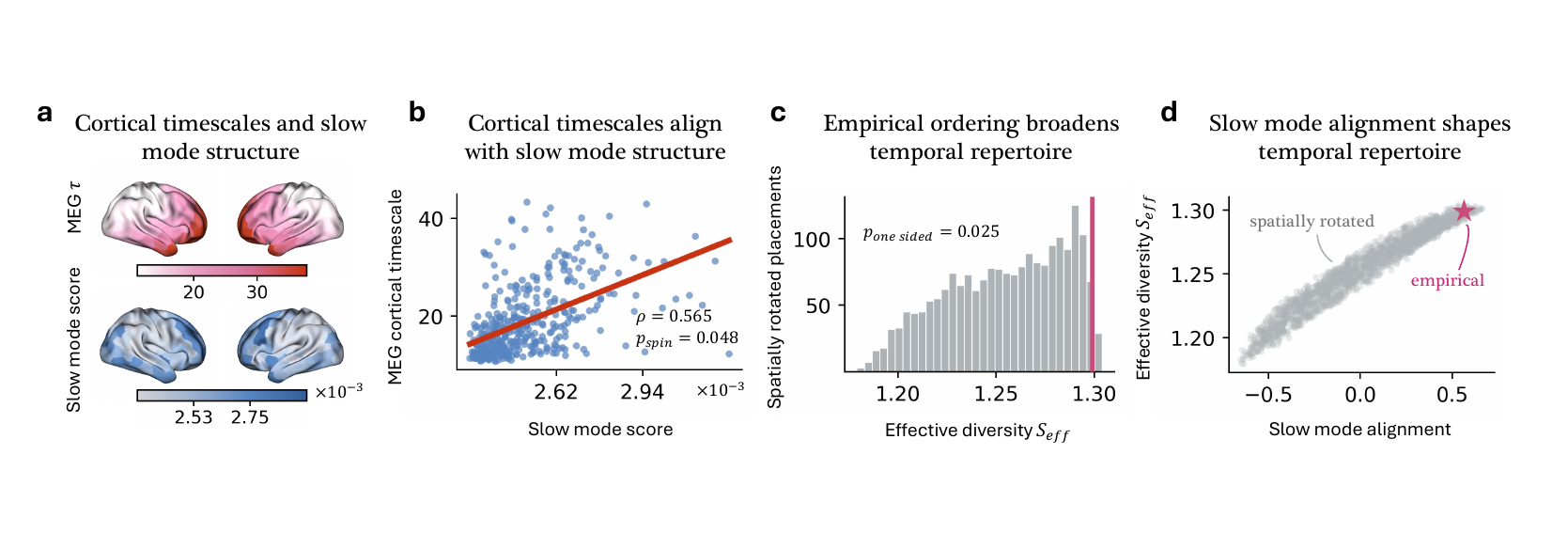}
    \caption{\textbf{Human cortical timescale organization is consistent with predicted dynamics of slow mode placement.}
    \textbf{a} Empirical MEG based cortical timescales and structural connectome slow mode score cortical mapping.
    \textbf{b} Cortical timescales increase with the slow mode score; significance is assessed with spatial rotations.
    \textbf{c} The empirical cortical ordering leads to broader effective temporal repertoire \(S_{\mathrm{eff}}\) compared to spatially rotated null placements.
    \textbf{d} Throughout rotated placements, stronger slow mode alignment is associated with a higher effective temporal diversity.}
    \label{fig:connectome}
\end{figure}

\revised{Finally, we examine whether measured cortical activity timescales are associated with slow recurrent modes of the human structural connectome.} The empirical MEG neural timescales were positively associated with the slow mode score of the structural connectome across 400 regions (\(\rho=0.565\), two-sided spatial rotation \(p=.048\); Fig.~\ref{fig:connectome}a,b). Regions with longer cortical timescales therefore participated more strongly in slowly decaying connectome modes. Importantly, the empirical cortical ordering exhibited the dynamical consequence predicted by network experiments. With the structural connectome and timescale set fixed, the empirical ordering produced \(S_{\mathrm{eff}}=1.299\), at the \(97.6\)th percentile among spatially rotated placements (\(p=.025\), one-sided; Fig.~\ref{fig:connectome}c). This indicates, in the connectome model, the spatial arrangement of cortical timescales contributes to the temporal repertoire beyond the timescale set itself. For rotated placements at \(g=0.90\), stronger slow mode alignment was associated with more diverse effective dynamics (Fig.~\ref{fig:connectome}d), in line with the relationship between placement and dynamics established in experiments. \revised{The dynamical result depends on assigning the measured MEG timescale ordering to intrinsic time constants in the connectome model. This remains a modelling assumption rather than a direct measurement of intrinsic neuronal time constants in the cortex.}

\revised{To examine whether recurrent connectivity itself could generate part of the empirical timescale ordering, we also tested a homogeneous model ($g=0.90$, $\tau_0=8$ for all 400 regions). The resulting effective timescales showed a positive association with MEG activity timescales (Spearman $\rho=0.523$, spatial rotation $p=.063$). Including slow mode participation as an additional predictor increased rank-based explained variance by $\Delta R^2=0.1069$ (spatial rotation $p=.020$), with similar increases for $\tau_0=4$ and $16$ (Appendix~\ref{app:connectome}).}

\section{Discussion}

Our study shows that rearranging an identical set of intrinsic time constants on a recurrent graph with fixed weights changes effective dynamics and computation. This indicates that effective temporal heterogeneity depends not only on intrinsic timescale diversity, but also on how these properties align with recurrent dynamics. Previous work established computational benefits of heterogeneous timescales, effects of spatially structured local heterogeneity, and relationships between inferred regional decay rates and slow network modes \citep{Tanaka2022,Demirtas2019,Kim2025}. \revised{Our contribution is a controlled manipulation of timescale placement, rather than a new cognitive task or RNN architecture.}

The comparison between structural and slow mode coordinates suggests that placement effects are better captured by distributed recurrent dynamics than local connectivity. Degree can carry meaningful dynamical information when heterogeneity gives rise to slow hub dynamics \citep{Zenari2026}, but its relevance depends on architecture. Cortical timescale hierarchies can likewise depend on local and long-range recurrence \citep{Chaudhuri2015,Stern2023}. Local graph measures may therefore transfer less reliably between architectures. In the tested leaky-ReLU setting, across topologies and tasks, slow mode alignment consistently shaped effective temporal dynamics and computation, without implying that the reference modes fully determine nonlinear dynamics.

The functional effects show that placement influences not only spontaneous network dynamics but also computation. With separately fitted readouts, performance differences reflect temporal information made linearly accessible from the recurrent state, and generally larger effects at longer delays link this organization to memory retention. Placement also biased early learning in MemoryPro, in line with evidence that pre-learning dynamics can shape subsequent learning \citep{Kurikawa2025}. \revised{Training recurrent weights partly compensated for unfavorable initial placement. Previous work has also shown that long timescales can arise through either local or recurrent mechanisms \citep{Khajehabdollahi2024}.} \revised{Matching mean log effective timescales and selectively perturbing slow and fast recurrent modes provide further evidence for the functional contribution of slow modes, although these controls do not establish mediation by timescale spread.} \revised{Thus, placement acts as an initial temporal bias that recurrent training can partly offset. Learning intrinsic time constants also improved MemoryPro performance without producing positive mean alignment with slow recurrent modes, indicating that task learning does not necessarily converge to aligned placement.}

\revised{The cortical findings are based on regional activity timescales rather than direct intrinsic neuronal constants \citep{Shafiei2023,Zeraati2026}. The association between MEG timescales and slow connectome modes is empirical, whereas the increased diversity of effective timescales in the connectome model depends on assigning MEG ranks to intrinsic time constants (Fig.~\ref{fig:connectome}). A homogeneous connectome model also generated effective timescales associated with MEG activity, with slow mode participation providing additional predictive information. These findings complement previous links between activity timescales and connectivity \citep{Fallon2020,Milisav2026}, and between inferred regional decay rates and slow modes \citep{Kim2025}. Measured activity timescales can vary with task demands \citep{Zeraati2023}, while local time constants can adapt in recurrent models \citep{Khajehabdollahi2024}; their interaction with placement remains untested.}

\revised{Our computational tests focused on the role of timescale placement in established cognitive tasks rather than complex sequence benchmarks. Fixed recurrence tests covered four tasks, three topologies, and 64--256 units; placement effects during recurrent training were tested only on MemoryPro. Future work should examine more complex tasks, larger networks, and biological measurements that distinguish local time constants from recurrence-generated activity timescales.}

\newpage
\subsection*{AI use statement}

In this work, we used generative AI tools to design or provide feedback on research methodology or experiments, implement methods, and interpret results. We have not used generative AI tools to propose or refine hypotheses, generate synthetic data sets, help develop theoretical models or conceptual frameworks, formulate mathematical claims, provide critical ingredients for proving mathematical claims, assist in the writing of proofs, assist with translation, clean and reformat dataset, or support qualitative and thematic data analysis. Additionally, we used generative AI tools to create or edit software code, identify relevant literature, and edit a research paper to improve readability. All AI-assisted work was reviewed by the authors. Literature identified with AI assistance was checked against the original sources before citation, and AI-generated or modified code was reviewed and tested. The authors made all final scientific and methodological decisions and take responsibility for the content of this work.

\subsection*{Reproducibility statement}

Definitions of the initial diagnostic tasks and training details are given in
Appendix~\ref{app:tasks}; the placement protocol and slow mode coordinate are
detailed in Appendices~\ref{app:placement}-\ref{app:slowmode}; the fixed
recurrence behavioral and learning experiments are described in
Appendices~\ref{app:taskedplacement} and~\ref{app:learning}; \revised{additional placement controls and learned time constant analyses are in Appendices~\ref{app:additional_controls} and~\ref{app:learnable_tau};} and the human
connectome analyses are detailed in Appendix~\ref{app:connectome}.

\bibliography{topodyn_4_2}
\bibliographystyle{preprint}

\appendix
\linespread{0.98}\selectfont

\section{Diagnostic tasks and training details}
\label{app:tasks}

\noindent\textbf{Diagnostic tasks}
The first experiment uses two diagnostic tasks, T1 and T2.
In T1, two independent latent components each follow a stationary AR(1) process,
\begin{equation}
z_k(t)=\rho_k z_k(t-1)+\sqrt{1-\rho_k^2}\,\epsilon_k(t),
\qquad
\rho_k=e^{-1/T_k},
\end{equation}
where \(\epsilon_k(t)\sim\mathcal{N}(0,1)\) and \(z_t=(z_1(t),z_2(t))^\top\).
The latent state \(z_t\) is not observed directly; the network receives
\begin{equation}
x_t=Mz_t+\eta_t,
\end{equation}
where \(M\) is orthogonal and \(\eta_t\sim\mathcal{N}(0,I_2)\) is independent
observation noise. The network is trained to reconstruct \(z_t\). We keep the
geometric mean of the two latent timescales fixed at \(8\) and vary their
separation, so that \(T_{\mathrm{fast}}=8/\sqrt{s}\) and
\(T_{\mathrm{slow}}=8\sqrt{s}\), where \(s\) is the timescale ratio.
We test \(s\in\{1,2,4,8,16\}\), and the comparison across model configurations
(non-recurrent leaky units, reservoir, and RNN) uses the hardest condition,
\(s=16\). Performance is evaluated after a \(32\)-step burn-in.

For T2, the latent signals are three independent white noise sequences,
\begin{equation}
u_t\sim\mathcal{N}(0,I_3),
\end{equation}
with \(u_t=(u_1(t),u_2(t),u_3(t))^\top\).
The network observes only the noisy orthogonal mixture,
\begin{equation}
x_t=Mu_t+\eta_t,
\qquad
\eta_t\sim\mathcal{N}(0,I_3),
\end{equation}
and is trained to recover each latent signal after a specified delay,
\begin{equation}
y_k(t)=u_k(t-d_k).
\end{equation}
Thus, T1 tests whether the network can unmix latent signals with different
timescales, and T2 tests whether it can retain multiple latent signals
over a delay. The comparison across model configurations uses
\((d_1,d_2,d_3)=(4,8,16)\), and T2 performance is evaluated after a burn-in
of \(\max_k d_k+16\) steps.

\noindent\textbf{Training and model configuration}
Throughout these diagnostic experiments, time is expressed in model steps
(\(\Delta t=1\)). Models here use \(64\) hidden units and sequence
length \(256\), and are trained for \(2000\) steps with batch size \(128\)
using Adam with learning rate \(10^{-3}\), no weight decay, and gradient norm
clipping at \(1\), with five paired model seeds. Homogeneous models learn one
globally shared \(\tau\), and heterogeneous models learn separate
\(\tau_i\) for each unit. Learned timescales are initialized around \(8\) and
constrained to \([0.5,128]\), and the checkpoint with the lowest validation
error is used for held-out evaluation. Training uses mean squared error and all
target channels are weighted equally.

The reservoir and RNN configurations use the same heavy-tail graph
construction with density \(0.15\), and the initial masked recurrent matrix is
rescaled to spectral radius \(0.9\). Recurrent weights are fixed in the
reservoir and optimized under the same fixed graph mask in the RNN; the readout
and intrinsic timescales are learned in both.

\noindent\textbf{Performance normalization}
Performance is normalized relative to a memoryless baseline and an analytic
causal reference,
\begin{equation}
G=
\frac{E_{\mathrm{base}}-E_{\mathrm{model}}}
     {E_{\mathrm{base}}-E_{\mathrm{ref}}},
\end{equation}
so that \(G=0\) corresponds to memoryless performance and \(G=1\) to the
analytic reference. For T1, \(E_{\mathrm{base}}\) is the error of the Bayes
estimate \(\mathbb{E}[z_t\mid x_t]\), which uses just the current observation,
and \(E_{\mathrm{ref}}\) is the error of the exact causal Kalman estimate
\(\mathbb{E}[z_t\mid x_{0:t}]\), which uses the full observation history.
For T2, \(E_{\mathrm{base}}\) is the error of predicting the prior mean of zero,
and \(E_{\mathrm{ref}}\) is the error of the minimum mean squared error estimate obtained from the
corresponding delayed observation. Heterogeneity effects are shown as
\(\Delta G=G_{\mathrm{heterogeneous}}-G_{\mathrm{homogeneous}}\).
Reported 95\% confidence intervals are percentile bootstrap intervals from
\(20{,}000\) resamples over the five paired model seeds.

\begin{figure}[!ht]
    \centering
    \includegraphics[
        width=\textwidth,
        trim=20 35 48 24,
        clip
    ]{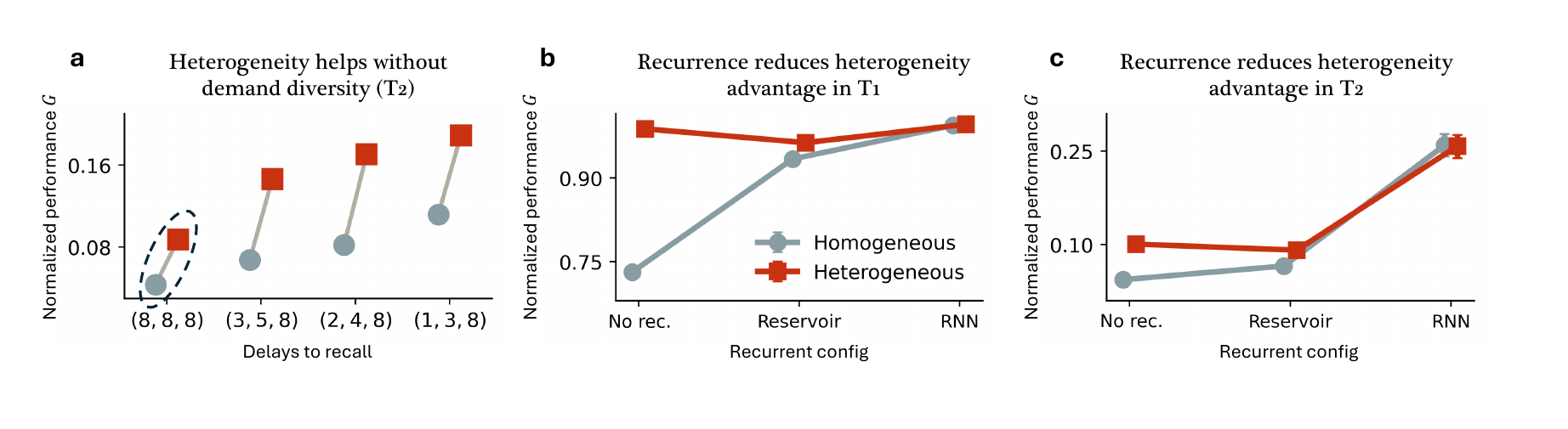}
    \caption{\textbf{Heterogeneity remains beneficial without demand diversity; recurrence reduces the heterogeneity advantage.}
    \textbf{a} T2 performance for different target delay combinations. Heterogeneous
    leaky units outperform homogeneous leaky units, including when all three channels have the same
    delay, \((8,8,8)\).
    \textbf{b} Normalized performance across leaky units, reservoir, and RNN
    configurations on T1 at \(s=16\).
    \textbf{c} The same comparison on T2 with
    \((d_1,d_2,d_3)=(4,8,16)\). In both tasks, homogeneous performance approaches
    heterogeneous performance with recurrence added.}
    \label{fig:supp_tasks}
\end{figure}

To test if the T2 heterogeneity advantage depends on having different
delay demands across channels, we hold the maximum delay at \(8\) and compare
\((8,8,8)\), \((3,5,8)\), \((2,4,8)\), and \((1,3,8)\).
Heterogeneous leaky units remain better than homogeneous leaky units even in
the equal delay condition (Fig.~\ref{fig:supp_tasks}a). Normalized performance
across model configurations shows that homogeneous performance approaches
heterogeneous performance with recurrence added in both T1 and T2
(Fig.~\ref{fig:supp_tasks}b,c).

\noindent\textbf{Temporal representation diagnostics}
We examine temporal representations in the trained leaky units in T2 with
\(d_1=d_2=d_3=8\). For a unit with intrinsic timescale \(\tau_i\), its temporal
kernel is
\begin{equation}
k_i(\ell)=\alpha_i(1-\alpha_i)^\ell,
\qquad
\alpha_i=1-e^{-1/\tau_i}.
\end{equation}
Collecting these kernels as columns of \(K\), we define the delay capacity
\(C_K(d)\) as how well their span represents an input at delay \(d\), obtained
by projecting the corresponding unit impulse \(e_d\),
\begin{equation}
\hat e_d=KK^{+}e_d,
\qquad
C_K(d)=\|\hat e_d\|_2^2 .
\end{equation}
Because \(\|e_d\|_2=1\), larger \(C_K(d)\) indicates better representation of
that delay.

We also quantify how much past input remains recoverable from the state
of the frozen trained network. Each trained model is driven by independent
Gaussian white noise \(r_t\sim\mathcal{N}(0,I_3)\). After a \(64\) step
burn-in, a readout is fitted to recover each
input channel at lags \(0\) through \(64\) and evaluated on independent
sequences. The input memory at lag \(d\) is
\begin{equation}
m(d)=\frac{1}{3}\sum_{c=1}^{3}
\operatorname{corr}^2\!\left(\hat r_{t-d,c},r_{t-d,c}\right).
\end{equation}
The readout is fitted on \(64\) sequences of length \(320\) and evaluated on
\(48\) independent sequences. Three white noise realizations are averaged
within each trained model, with the five paired model seeds serving as the
unit of replication. Thus, the delay capacity \(C_K(d)\) measures delay
representation by the learned temporal kernels, and the input memory
\(m(d)\) measures recoverable information about past inputs.

\noindent\textbf{Effective timescale diagnostic}
For the effective timescale comparison in Fig.~\ref{fig:main}d, the T2
\((4,8,16)\) checkpoints from each model configuration are driven by the same
three Gaussian white noise input batches of \(48\) sequences each. After the
\(32\) step T2 burn-in, each unit's effective timescale
\(\tau_i^{\mathrm{net}}\) is estimated from an exponential fit to its hidden
state autocorrelation. Per-unit estimates are geometrically averaged among the
input batches before computing
\(S_{\mathrm{eff}}=\operatorname{std}_i[\log\tau_i^{\mathrm{net}}]\).

\section{Placement experiment details}
\label{app:placement}

Placement experiments use \(64\) unit reservoirs with fixed recurrent weights
and step time units (\(\Delta t=1\)). Erd\H{o}s-R\'enyi (ER) and
heavy-tail graphs have exactly the same off-diagonal edge budget,
corresponding to density \(0.15\). ER edges are sampled uniformly, and
heavy-tail graphs use a bounded heterogeneous degree distribution. 
Recurrent weights are obtained by masking an
orthogonally initialized matrix and rescaling the masked recurrent
matrix to spectral radius \(0.9\). We use \(20\) independently generated graph
realizations and two recurrent weight initializations per graph, which gives \(40\)
fixed recurrent network instances per topology. Every instance receives the same
fixed set of \(64\) intrinsic timescales, log-spaced from \(1\) to \(64\).

Placement alignment is quantified by Spearman correlation between intrinsic
timescale and a node coordinate (degree, strength, or slow mode
coordinate; App.~\ref{app:slowmode}). We test target alignments
\(-1,\;-2/3,\;-1/3,\;0,\;1/3,\;2/3,\;1\).
At the aligned regime (\(\rho_s=1\)), the slowest intrinsic timescales are assigned to the
highest scoring nodes; the assignment is reversed at the anti-aligned regime.
Intermediate placements are generated by permuting the same timescale set and,
from \(256\) candidate permutations, selecting the assignment whose achieved
Spearman correlation is closest to the target. Two placement replicates are
evaluated at each alignment level. Placement construction depends only on the
fixed node coordinate and achieved alignment; effective dynamics are not used
to select an assignment.

After each placement, the reservoir is driven by the same fixed four channel
Gaussian white noise. This contains \(8\) sequences of length
\(384\), with a \(96\) step burn-in. Effective timescales are
estimated from unit activity autocorrelations using lags up to \(64\), and the resulting temporal repertoire is
quantified by \(S_{\mathrm{eff}}\) as defined in the main text. Thus, graph,
recurrent weights, intrinsic timescale set, and external drive are all fixed
within each network instance and only the nodal assignment of timescales changes.
Confidence intervals for the placement effects use \(4000\) hierarchical
bootstrap resamples over graph realizations and recurrent weight
initializations.

\section{Slow mode coordinate}
\label{app:slowmode}

We compare two local structural coordinates with a recurrent dynamics based
coordinate. The structural coordinates are off-diagonal binary degree and
off-diagonal in-strength (sum of absolute incoming weights). Motivated by the
eigenvector localization view of \citet{Chaudhuri2014}, the slow mode
coordinate quantifies where slowly decaying recurrent modes are expressed
across nodes.

For each fixed recurrent substrate, we construct a homogeneous reference system
in which every unit has the same intrinsic timescale \(\tau_0=8\). The
reference system retains the recurrent substrate but replaces the heterogeneous
timescale assignment by this common value. It is driven by a fixed
four channel Gaussian white noise reference input that is independent of the
input used to measure \(S_{\mathrm{eff}}\). Using \(8\) sequences of length
\(384\) and a \(96\) step burn-in, we estimate how often each ReLU unit is
active and form the reference Jacobian
\begin{equation}
J_0=(1-\alpha_0)I+\alpha_0\bar{D}W_{\mathrm{rec}},
\qquad
\alpha_0=1-e^{-1/\tau_0},
\end{equation}
where \(\bar{D}_{ii}\) is the fraction of post burn-in samples for which unit
\(i\) has a positive ReLU preactivation.

We then decompose \(J_0\) into recurrent modes. Modes with eigenvalues closer
to the unit circle decay more slowly. Each node is assigned a score from its
normalized squared loading on each right eigenvector, with slower modes
receiving greater weight,
\begin{equation}
s_i=\sum_k w_k
\frac{|v_{ik}|^2}{\|v_k\|_2^2},
\qquad
w_k\propto-\frac{1}{\log|\lambda_k|},
\end{equation}
where \(\lambda_k\) and \(v_k\) are the eigenvalue and right eigenvector of mode
\(k\). The resulting slow mode coordinate measures the spatial expression of
slow recurrent modes across nodes. It is computed once from the homogeneous
reference system before heterogeneous intrinsic timescales are assigned and is
then kept fixed across all tested placements.

\section{Tasked placement across temporal demand}
\label{app:taskedplacement}

\noindent\textbf{Network construction and calibration}
The fixed recurrence behavioral study uses 128-unit leaky-ReLU reservoirs with
\(\Delta t=20\) ms. All topology families have the same off-diagonal edge
budget and density \(0.15\). ER and heavy-tail graph constructions are as above;
modular graphs contain four modules, with \(0.10\) of off-diagonal edges assigned
between modules while preserving the same total edge count. Recurrent weights are obtained by masking an
orthogonally initialized matrix with the graph and rescaling the resulting
matrix to spectral radius \(0.9\). Input weights use Xavier initialization and
biases are zero. Every network receives the same fixed set of 128 intrinsic
timescales, log-spaced from \(40\) to \(640\) ms with geometric mean \(160\) ms.

Calibration and confirmation were separated. The intrinsic timescale center and
span were selected using behavioral viability under random placement. Ridge
regularization and task difficulty were also calibrated using random placements
only. All timescale, readout, and task settings were frozen before
the aligned versus anti-aligned confirmation analyses.

\noindent\textbf{Cognitive tasks}
MemoryPro and MemoryAnti use continuous circular stimuli and responses following delay-go and delay-anti task families used in multitask RNN studies
\citep{Yang2019,Driscoll2024}. In MemoryPro, the network maintains a stimulus
direction over a delay and responds in that direction; in MemoryAnti,
the response is displaced by \(\pi\). We test delays of
\(200, 400, 800, 1200, 1600, 2000,\) and \(2400\) ms. Performance is measured
as mean angular error.

Perceptual decision making (PDM) requires integrating noisy binary evidence,
maintaining the resulting decision through a variable delay, and reporting a
binary choice \citep{Inagaki2019}. Each trial contains a \(1200\) ms evidence
period followed by a delay of \(0, 200, 400, 800, 1200, 1600,\) or \(2400\) ms.
DelayMatchCategory requires retaining the category of a sample direction over a
delay and reporting whether a subsequent test direction belongs to the same
category \citep{FreedmanAssad2006}. Sample and test directions are drawn relative
to two circular categories, with delays of \(200, 400, 800, 1200, 1600, 2000,\)
and \(2400\) ms. Performance on both binary tasks is reported as percentage
error, \(100(1-\mathrm{balanced\ accuracy})\).

For each network, latent task variables and stochastic input streams are generated once and reused across placement conditions. Demand levels are paired; longer delays thus extend the same stochastic input stream instead of introducing independent noise. A separate ridge readout is fitted for each placement using the same regularization and matched training data, and evaluated on matched held-out trials. The graph, recurrent and input weights, biases, timescale set, trial variables, and input streams are otherwise fixed.

\noindent\textbf{Placement confirmation and inference}
Confirmation uses 16 independently generated networks from each of the ER,
heavy-tail, and modular families. For each network, placement is varied with
respect to either the slow mode coordinate or degree. We test seven target
Spearman alignments,
\(-1,\;-2/3,\;-1/3,\;0,\;1/3,\;2/3,\;1\).
The aligned and anti-aligned placements use the exact rank ordering of the
timescale set. At each intermediate alignment, three independently generated
permutations are selected to lie within \(0.03\) of the target correlation and
averaged before inference.

The primary placement contrast is
\begin{equation}
B(d)=E_{\mathrm{anti}}(d)-E_{\mathrm{aligned}}(d),
\end{equation}
where \(E\) is angular error for MemoryPro and MemoryAnti or percentage error
for PDM and DelayMatchCategory, so \(B(d)>0\) indicates an aligned placement
benefit. Means and 95\% confidence
intervals use \(5000\) paired bootstrap resamples over networks, and significance
is assessed with paired sign-flip tests. Effects are reported separately
for each topology family.

\begin{figure}[!htbp]
    \centering
    \includegraphics[
        width=\textwidth,
        trim=21 7 36 8,
        clip
    ]{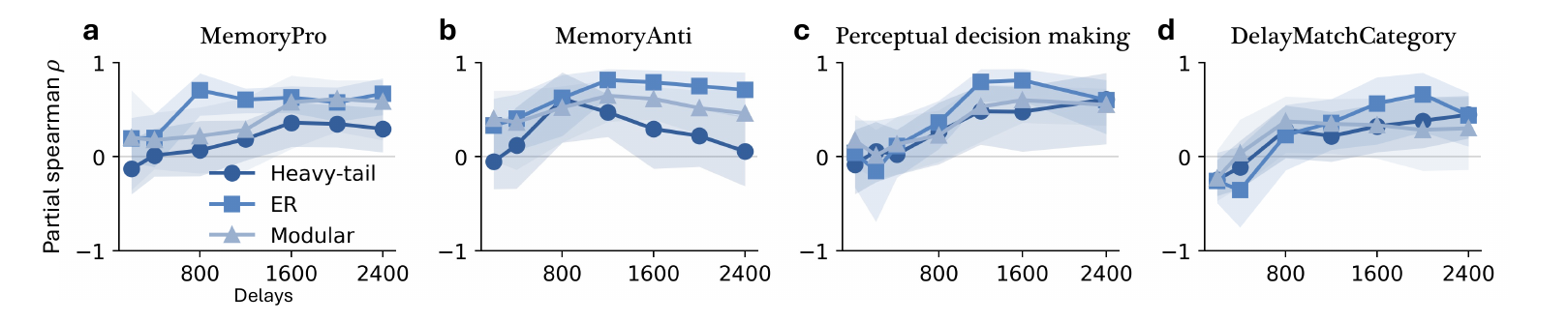}
    \caption{\textbf{Association between effective temporal repertoire and behavioral performance across temporal demand.}
    Partial Spearman correlations between \(S_{\mathrm{eff}}\) and task
    performance across the seven slow mode alignment levels for MemoryPro,
    MemoryAnti, PDM, and DelayMatchCategory,
    respectively, controlling for mean effective
    network timescale.}
    \label{fig:supp_seff_behavior}
\end{figure}

\noindent\textbf{Effective temporal repertoire and behavioral benefit}
For each network and demand level, we computed the partial Spearman correlation
between \(S_{\mathrm{eff}}\) and task performance across the seven slow mode
alignment levels, controlling for mean effective network timescale.
Network-level correlations were summarized within each topology family, with
95\% confidence intervals obtained by bootstrap resampling over networks. The
full demand-dependent trajectories are shown in Fig.~\ref{fig:supp_seff_behavior}.

\begin{figure}[!htbp]
    \centering
    \includegraphics[
        width=\textwidth,
        trim=21 7 36 8,
        clip
    ]{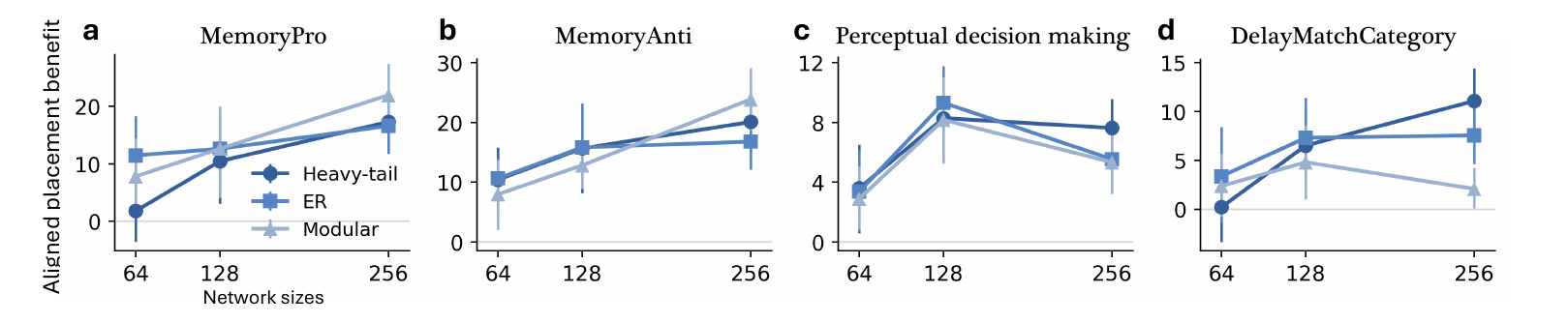}
    \caption{\textbf{Slow mode placement benefit across network size.}
    Aligned placement benefit \(B(1600)\) for 64, 128, and 256 unit networks
    across MemoryPro, MemoryAnti, PDM, and
    DelayMatchCategory. Points show mean paired placement effects across
    networks and error bars show 95\% bootstrap confidence intervals.}
    \label{fig:supp_size_robustness}
\end{figure}

\noindent\textbf{Network size robustness}
We tested whether the slow mode placement effect depended on reservoir size using
64, 128, and 256 unit networks, with the network construction and task, readout,
and placement protocols unchanged. Size robustness was evaluated at
\(1600\) ms rather than the maximal \(2400\) ms delay because the 64-unit
networks approached chance performance at \(2400\) ms under both placements. Aligned placement produced a positive behavioral benefit in all 36 task,
topology, and network size combinations (Fig.~\ref{fig:supp_size_robustness}).
Confidence intervals were entirely above zero in 8 of 12 comparisons for
64-unit networks and in all comparisons for 128 and 256 unit networks,
suggesting that the placement effect was not specific to the 128-unit setting.

\section{Additional controls for timescale placement and recurrent modes}
\label{app:additional_controls}

\revised{\noindent\textbf{Weighted eigenvector centrality baseline}
We evaluated 24 independently generated 128-unit reservoirs (eight each from Erd\H{o}s--R\'enyi, heavy-tail, and modular networks) with the same intrinsic timescale set, recurrent weights, and PDM evaluation for the paired placement comparisons. Weighted eigenvector centrality was computed from the absolute recurrent weight matrix. For each coordinate, aligned and anti-aligned timescale assignments were compared within the same reservoir. The benefit is $B(d)=E_{\mathrm{anti}}(d)-E_{\mathrm{aligned}}(d)$, where $E$ is the PDM error in percentage points. The mean benefits at 1600 ms were 8.34 points for slow-mode placement and 3.15 points for weighted eigenvector centrality, with a paired difference of 5.19 [2.76, 7.47], $p<.001$. At 2400 ms, the paired difference was 2.59 [0.13, 4.91], $p=.003$. The increase in effective timescale diversity was also greater for slow-mode placement (mean difference in $\Delta S_{\mathrm{eff}}=0.0593$; positive in 23 of 24 networks).}

\revised{\noindent\textbf{Matched mean effective timescale control}
In a separate set of 24 reservoirs, we sought placements with different effective timescale spreads but similar mean log effective timescales. Of these, 22 met the matching criterion and two were excluded. We independently checked the matching and found that the median absolute gap in mean log effective timescale was 0.0153 (maximum 0.0693), below the prespecified tolerance of 0.120. All 22 pairs had greater $S_{\mathrm{eff}}$ in the intended higher-spread placement (median difference 0.1150; minimum 0.0688). The paired PDM error advantage of the higher-spread placement is reported in Table~\ref{tab:additional_controls}.}

\begin{table}[!ht]
\centering
\small
\caption{Additional functional controls. Positive differences in the first two rows indicate lower PDM error for the higher-spread placement at matched mean log effective timescale. Positive differences in the last two rows indicate greater loss of the aligned benefit under slow rather than fast mode perturbation. All differences and confidence intervals are in PDM error percentage points.}
\label{tab:additional_controls}
\vspace{6pt}
\begin{tabular}{lcccc}
\hline
Control & Delay & Difference (pp) & 95\% CI & $p$ \\
\hline
Matched mean log effective time & 1600 ms & 7.53 & [5.48, 9.44] & 0.00005 \\
Matched mean log effective time & 2400 ms & 4.42 & [2.89, 6.02] & 0.00005 \\
Slow vs. fast perturbation & 1600 ms & 6.47 & [4.21, 8.82] & 0.00005 \\
Slow vs. fast perturbation & 2400 ms & 2.57 & [0.59, 4.41] & 0.00780 \\
\hline
\end{tabular}
\end{table}

\revised{\noindent\textbf{Slow versus fast recurrent mode perturbations}
We measured the effect of perturbing slow or fast recurrent subspaces during the memory delay in 24 independently generated reservoirs. Perturbation strengths were adjusted using task-independent Gaussian inputs to remove comparable amounts of network activity (median relative removed-energy difference 1.85\%; maximum 4.20\%). The slow and fast subspaces showed some overlap (median 0.325). Table~\ref{tab:additional_controls} reports the difference in loss of the aligned versus anti-aligned placement advantage under slow and fast perturbations.}

\section{Learning intrinsic time constants}
\label{app:learnable_tau}

\revised{We also examined whether intrinsic time constants learned during task optimization align with slow recurrent modes. Nine independently generated networks (three from each topology family) were tested on MemoryPro at 2400 ms with fixed or trainable intrinsic time constants. We made this comparison separately for fixed and trainable recurrent weights. Late error was measured as normalized angular error at the final training checkpoint. Learning the time constants reduced late error by 0.1365 on average with fixed recurrence ($p=.004$). With trainable recurrence, learning the time constants reduced error by an additional 0.00109 relative to fixed time constants ($p=.043$). The mean final Spearman correlation between learned time constants and slow recurrent modes was $-0.017$ with fixed recurrence and $-0.027$ with trainable recurrence. These results show that optimizing task performance does not necessarily lead to alignment of learned time constants with slow recurrent modes.}

\section{Learning, recurrent plasticity, and dynamical reorganization}
\label{app:learning}

\noindent\textbf{Experimental design and placement}
The learning study uses the same 128-unit network construction, topology families, timestep, and 40-640 ms intrinsic timescale set as the fixed recurrence task study. We generated eight independent networks from each topology family (24 networks total) and trained MemoryPro at delays of 400, 800, 1600, and 2400 ms. Each network is evaluated with two placement rules (slow mode and binary degree), two target alignments (\(-0.5\) and \(+0.5\)), and two recurrence conditions (fixed and trainable), which gives 768 runs. A placement is constructed once for each network, rule, and sign and reused across all four delays.

Slow mode placements target the prescribed Spearman correlation while constraining absolute correlations with degree and weighted in-strength to at most 0.10, along with a rank quintile balance constraint. Degree placements target the same \(\pm0.5\) correlations with degree but are not constrained with respect to the slow mode coordinate, as the two coordinates covary in some topology families. Achieved correlations with slow mode, degree, and weighted in-strength are retained for the secondary coordinate analysis. Placement construction depends only on the recurrent network and fixed intrinsic timescales, not on task performance.

\noindent\textbf{Training and pairing}
The output readout is trained in every condition. In trainable recurrence, recurrent weights are optimized along with the readout (retaining the original graph mask); in fixed recurrence, recurrent weights are unchanged. No spectral radius projection or recurrent gain renormalization is applied during learning. Optimization uses Adam for 2000 steps with batch size 32, recurrent learning rate \(3\times10^{-4}\), readout learning rate \(3\times10^{-2}\), no weight decay, and gradient norm clipping at 100. Performance is evaluated on 160 held-out trials at steps 0, 50, 100, 200, 300, 400, 600, 800, 1200, 1600, and 2000. Within each network and delay, compared placement and recurrence conditions use matched readout initialization, training batches, and evaluation trials.

\noindent\textbf{Learning efficiency and coordinate analysis}
Early learning efficiency is quantified from held-out error over the first 600 optimization steps. Baseline adjusted early learning gain \(A\) summarizes the average reduction in held-out error from its pre-training value across this window, integrated over the recorded checkpoints. The first 600 steps capture the early, pre-convergence phase of learning; larger values indicate faster improvement. For the primary analysis, we consider slow mode placements with trainable recurrence. Within each network and delay, \(\Delta A(d)=A_{+0.5}(d)-A_{-0.5}(d)\). A least-squares slope of \(\Delta A\) against \(\log_2 d\) is computed across the four delays within each network, so each network contributes one estimate of how the aligned learning advantage changes with temporal demand.

We perform a secondary analysis using the achieved slow mode and degree correlations across all trainable recurrence conditions. For each of the 96 conditions (24 networks \(\times\) four rule/sign combinations), we compute the slope of early learning gain \(A\) across \(\log_2 d\), and fit
\begin{equation}
S_{na}=\gamma_n+\beta_{\mathrm{slow}}\rho^{\mathrm{slow}}_{na}
+\beta_{\mathrm{deg}}\rho^{\mathrm{deg}}_{na}+\epsilon_{na},
\end{equation}
where \(S_{na}\) is the demand slope for condition \(a\) in network \(n\), \(\gamma_n\) is a network fixed effect, and the predictors are the achieved slow mode and degree correlations. Confidence intervals and two-sided tests use 5000 topology stratified cluster bootstrap resamples over networks. As a topology sensitivity analysis, we refit the model after omitting each topology family in turn. The slow mode coefficient remained positive when omitting heavy-tail (\(\beta_{\mathrm{slow}}=0.045\,[0.018,0.072]\)), modular (\(0.045\,[0.014,0.072]\)), or ER networks (\(0.059\,[0.031,0.088]\)).

\noindent\textbf{Plastic compensation and recurrent reorganization}
Compensation is assessed at the maximal 2400 ms delay using slow mode placement. Late error is the mean held-out error over checkpoints 1200, 1600, and 2000. Within each network, we compare the aligned versus anti-aligned late performance gap under fixed and trainable recurrence; a smaller gap with trainable recurrence indicates compensation for unfavorable initial placement.

For monitoring the evolving slow recurrent structure during learning, we construct a homogeneous timescale reference Jacobian from the current \(W_{\mathrm{rec}}\) at steps 0, 400, 800, 1200, 1600, and 2000. For longitudinal comparison, we track the slow invariant subspace rather than individual modes, avoiding the need to match modes across training checkpoints. Complex Schur decomposition is used to identify the invariant subspace associated with the 20\% of recurrent modes having the largest eigenvalue magnitudes. Each node is scored by its mean squared participation in this slow subspace, and its Spearman correlation with the fixed heterogeneous \(\tau_i\) assignment defines slow subspace alignment. In initially anti-aligned networks at 2400 ms, reorganization is quantified as the change in slow subspace alignment from step 0 to step 2000. This quantity is unchanged in fixed recurrence by construction, providing a reference for the trainable recurrence condition. Confidence intervals are obtained from 5000 bootstrap resamples over networks, and directional comparisons use one-sided Wilcoxon signed-rank tests.

\section{Human connectome analysis details}
\label{app:connectome}

\noindent\textbf{Data and connectome construction}
The structural connectome and MEG timescale map were obtained from the processed data release from \citet{Milisav2026}. The connectome was derived from diffusion-weighted MRI in 327 unrelated Human Connectome Project (HCP) participants using the HCP minimal preprocessing pipeline \citep{Glasser2013} and tractography-based structural network reconstruction; further details are described in \citet{Park2021}. Networks were represented in the 400-parcel Schaefer atlas \citep{Schaefer2018}. A distance-dependent consensus procedure was used to construct the group network, preserving mean density and the edge length distribution \citep{Betzel2019}; retained edge weights were averaged across participants, log transformed, and rescaled to \([0,1]\) \citep{Milisav2026}. The group-average MEG timescale map was from 33 unrelated HCP participants \citep{Shafiei2023} and was parcellated to the same Schaefer atlas by \citet{Milisav2026}. The structural and MEG data are therefore group-level representations, not subject-matched measurements.

\noindent\textbf{Slow mode score and spatial association}
The connectome is spectrally normalized with gain \(g=0.9\), and the homogeneous reference timescale is fixed at \(\tau_0=8\). For the reference system, the eigenvalues are
\begin{equation}
\mu_k=(1-\alpha_0)+\alpha_0\lambda_k(W),
\end{equation}
with mode decay times
\begin{equation}
T_k=-\frac{1}{\log|\mu_k|}.
\end{equation}
The parcelwise slow mode score is
\begin{equation}
s_i=\sum_k q_kv_{ik}^2,
\qquad
q_k=\frac{T_k}{\sum_jT_j},
\end{equation}
where \(v_{ik}\) is the loading of parcel \(i\) on mode \(k\). Association between the MEG timescale map and \(s_i\) is quantified using Spearman correlation. Spatial significance is assessed using \(10{,}000\) spherical spin permutations of the cortical map, which preserve its spatial autocorrelation \citep{AlexanderBloch2018}. The two-sided spatial \(p\)-value is computed from the absolute null correlations using a \(+1\) permutation correction.

\noindent\textbf{Dynamical placement analysis}
For the dynamical placement analysis, only the rank ordering of the empirical MEG map is retained and mapped onto a fixed set of 400 model timescales log-spaced from \(1\) to \(64\). We use \(2{,}000\) of the same spatial rotations to define alternative placements, so that the structural connectome and complete timescale set remain fixed and only their cortical assignment changes. For each placement,
\begin{equation}
J=I-D_\alpha+D_\alpha W,
\end{equation}
where \(D_\alpha=\operatorname{diag}(\alpha_i)\) and \(\alpha_i=1-e^{-1/\tau_i}\). Under Gaussian white noise input, stationary node autocorrelations are computed analytically for this linear system, and effective timescales are estimated from these autocorrelations using lags up to \(64\). The resulting temporal repertoire is quantified by \(S_{\mathrm{eff}}\) as defined in the main text. The empirical ordering is evaluated relative to the rotated null using its percentile and a one-sided upper tail permutation test with a \(+1\) correction. Across rotated placements, slow mode alignment is quantified by the Spearman correlation between assigned model timescales and the slow mode score, and its association with \(S_{\mathrm{eff}}\) is assessed using Spearman correlation.

\revised{\noindent\textbf{Homogeneous recurrence control}
To examine the extent to which the recurrent connections within the connectome account for part of the observed spatial variability in measured activity timescales, we used the empirical 400-region connectome and assigned the same reference intrinsic time constant $\tau_0$ to every region, with gain fixed at $g=0.90$. The model's generated effective timescales were compared with the empirical group-average MEG activity timescale map using Spearman correlation and a two-sided spatial rotation test. At $\tau_0=8$, the association was $\rho=0.523$ (spatial rotation $p=.063$). We further assessed whether parcelwise slow mode participation provides predictive information about MEG timescale ranks after accounting for the homogeneous-model effective timescale ranks. The additional rank-based explained variance was $\Delta R^2=0.1069$, with upper-tail spatial rotation $p=.020$. The increment was similar for reference time constants $\tau_0=4$ ($\Delta R^2=0.1067$) and $\tau_0=16$ ($\Delta R^2=0.1021$).}
\end{document}